\RequirePackage[2020-02-02]{latexrelease}
\documentclass[reprint,superscriptaddress,amsmath,amsfonts,amssymb,aps,prl,showkeys]{revtex4}

\usepackage{physics}
\usepackage[normalem]{ulem}
\usepackage{bm}
\usepackage{physics}
\usepackage{color}
\usepackage{graphicx}
\usepackage{hyperref}
\hypersetup{
  breaklinks = {true},
  citecolor = {blue},
  colorlinks = {true},
  linkcolor = {red},
}
\newcommand{\BB}{\mathcal{B}}

\begin{document}

\title{Electrical Control of Spin-polarized Topological Currents in Monolayer WTe$_2$ }
\author{Jose H. Garcia}
\affiliation{Catalan Institute of Nanoscience and Nanotechnology - ICN2, (CSIC and BIST), Campus UAB, Bellaterra, 08193 Barcelona, Spain}

\author{Jinxuan You}
\affiliation{Catalan Institute of Nanoscience and Nanotechnology - ICN2, (CSIC and BIST), Campus UAB, Bellaterra, 08193 Barcelona, Spain}
\affiliation{Simune Atomistics S.L., Tolosa Hiribidea, 76, 20018 Donostia-San Sebastian, Spain}
\affiliation{Department of Materials Science, Universitat Autònoma de Barcelona, Cerdanyola del Vallès, 08193 Bellaterra, Spain}
\author{M\'onica García-Mota}
\affiliation{Simune Atomistics S.L., Tolosa Hiribidea, 76, 20018 Donostia-San Sebastian, Spain}

\author{Peter Koval}
\affiliation{Simune Atomistics S.L., Tolosa Hiribidea, 76, 20018 Donostia-San Sebastian, Spain}

\author{Pablo Ordej\'on}
\affiliation{Catalan Institute of Nanoscience and Nanotechnology - ICN2, (CSIC and BIST), Campus UAB, Bellaterra, 08193 Barcelona, Spain}

\author{Ram\'on Cuadrado}
\affiliation{Catalan Institute of Nanoscience and Nanotechnology - ICN2, (CSIC and BIST), Campus UAB, Bellaterra, 08193 Barcelona, Spain}

\author{Matthieu J. Verstraete}
\affiliation{nanomat/QMAT/CESAM and European Theoretical Spectroscopy Facility
Universite de Liege, Allee du 6 Aout 19 (B5a), 4000 Liege, Belgium}

\author{Zeila Zanolli}
\affiliation{Dept. of Chemistry, Debye Institute for Nanomaterials Science, and ETSF, Utrecht University, The Netherlands}

\author{Stephan Roche}
\affiliation{Catalan Institute of Nanoscience and Nanotechnology - ICN2, (CSIC and BIST), Campus UAB, Bellaterra, 08193 Barcelona, Spain}
\affiliation{ICREA--Instituci\'o Catalana de Recerca i Estudis Avan\c{c}ats, 08010 Barcelona, Spain}

\date{\today}

\begin{abstract}
We evidence the possibility for coherent electrical manipulation of the spin orientation of topologically protected edge states in a low-symmetry quantum spin Hall insulator. By using a combination of {\it ab-initio} simulations, symmetry-based modeling, and large-scale calculations of the spin Hall conductivity, it is shown that small electric fields can efficiently vary the spin textures of edge currents in monolayer 1T'-WTe$_2$ by up to a 90-degree spin rotation, without jeopardizing their topological character. These findings suggest a new kind of gate-controllable spin-based device, topologically protected against disorder and of relevance for the development of topological spintronics. 
\end{abstract}

\maketitle

{\it Introduction.}
The existence of the quantum spin Hall (QSH) insulator \cite{Fei2017,Wu76} has boosted opportunities for spintronics and quantum metrology, given the ability of topologically protected states to convey spin information over long distances at ultralow dissipation rate \cite{HasanRMP2010, Sinova2015, Yang2016a}. QSH is a manifestation of strong spin-orbit coupling which fundamentally depends on the symmetries of the system \cite{Kane2005, Kane2005QSH,Bernevig1757, Bernevig2006}. However, even in time-reversal symmetric systems, the lack of a spin conservation axis in QSH insulators allows backscattering effects for edge states, limiting their ballistic transport \cite{Sheng2006,Strom2010,Schmidt2012}.  In some situations, the emergence of a phenomenon known as persistent spin texture (PST) enforces spin conservation and favors long spin lifetimes even in the presence chemical disorder and structural imperfections \cite{Schliemann2017}. Such an effect is deeply rooted in the underlying symmetries of the system \cite{Tao2018} and opens promising prospects for spintronics when combined with the manifestation of dissipationless chiral edge states. 

As a matter of fact, the recent prediction \cite{PRLGarcia2020} and experimental observations \cite{ZhaoCobden2020,ChengTan2020} of a PST-driven {\it canted quantum spin Hall effect} in low-symmetry monolayer WTe$_2$ provide new ingredient for the use of topological materials in spintronic applications \cite{Giustino_2020,Sierra2021}. Such phenomenon is rooted in the lack of multiple vertical mirror planes enabling a constant spin-texture in the Fermi level vicinity \cite{PRLGarcia2020}. The topologically protected edge-states inherit such canted spin polarization from the bulk bands leading to a quantized spin Hall conductivity (SHC) plateau of $2e^2/h$ along the canting axis. Notably, such state is more robust against inversion symmetry breaking when compared with higher symmetry systems (such as graphene), in which broken inversion symmetry usually generates a Rashba spin-orbit coupling (SOC) effects, inducing momentum-dependent spin textures and hindering spin conservation \cite{Min2006}. The Rashba SOC is more generally a consequence of uncompensated electric fields in noncentrosymmetric systems either originated from the substrate, strain, or the crystal geometry \cite{xia2012semiconductor}. Additionally, electrostatic gates larger than 1 V/nm typically lead to appreciable modulation of the Rashba SOC strength \cite{NittaPRL,Bindel2016,Premasiri2018}. Actually, transition metal dichalcogenides in the 1T' phase have already shown a variety of tunable properties under electric fields  \cite{Wang2016e,Bindel2016,Zhang2017,Xu2018,Maximenko2020}, resulting from spin splitting and modulation of the SOC parameters \cite{ShiLikun}. However a possible control of the canted QSH phase via electric fields remains unexplored to date, albeit it could enable the design of all-electrically controlled spin devices, such as spin-dependent topological switches, that would enrich the prospects for dissipationaless spintronics and quantum metrology. 

This Letter reports on the possibility of a fully controllable variation of up to 90 degrees rotation of the spin polarization of chiral edge-states, dictating the canted QSH effect, while preserving spin conservation. By combining density functional theory (DFT) with tight-binding methods and quantum transport simulations, we show that the emerging PST can be continuously varied from in-plane to out-of-plane under electric fields below 0.1 V/nm, making this effect experimentally accessible. The experimental confirmation of such fully electrically tunable spin-polarized topological currents would establish a new milestone towards replacing magnetic components in spintronic devices and all-electric spin circuit architectures, as well as optimized resistance quantum standards. 

{\it Model and electronic properties.} The relevant spin transport properties of monolayer WTe$_2$ in its 1T' and 1T$_{\rm d}$ structural phases originate from the presence of two charge pockets, $Q$ and $-Q$, symmetrically located around the $\Gamma$ point and related  by time-reversal symmetry. Those pockets have fundamentally $p-d$ hybridized orbital symmetry that can be well captured by a 4-band Hamiltonian \cite{PRLGarcia2020, Xu2018, Shi2019prb,Ok2019}. Ref.\cite{PRLGarcia2020} demonstrated the 1T$_{\rm d}$ phase conserves spin along a direction given by a canted angle $\varphi$ prescribed by the SOC parameters. Since we are focused on the physics around the $Q$ points, we shifted our Brillouin zone origin to these points, and through this process  our model becomes a tilted massive Dirac Hamiltonian \cite{Goerbig2008,Huang2016}
\begin{equation}
{\cal H} = \hbar\, v_{\rm F}^x\,( \tau_z  \sigma_x q_x + \zeta\sigma_y q_y) + \hbar{\bm{v}}_T\cdot \bm{q}  +\bm{\sigma}\cdot\bm{\BB}+ {\cal H}_{q^2}+{\cal H}_{\rm soc},
\end{equation}
where $\bm{q}=\bm{k}-\tau_z\bm{Q}$ the moment measured from the $Q$ points;  $\sigma_i$ with $i=x,y,z$ the Pauli matrices acting on the pseudospin space,  $v_F^x$ the Fermi velocity of electrons traveling along $y$ axis close to the $Q$ points, ${\zeta}$ an adimensional parameter that produces anisotropy in the Fermi velocity, ${\bm \BB}$ a Zeeman-like field acting on the pseudospin; $\bm{v}_T$ tilt velocity vector pointing to  $\bm{Q}$ and responsible for the warping of the Fermi contour, ${\cal H}_{q^2} =\bm{q}^2(1/m_p-\sigma_x/m_d)$ a quadratic term that describes the behavior of the electrons in the proximity of the $\Gamma$ point. The SOC term is given by
\begin{equation}
{\cal H}_{\rm soc} =  \tau_{z}\sigma_z ( \lambda_y s_y  + \lambda_z s_z )  + (\Upsilon s  \times \bm{q})\cdot \bm{\sigma},
\end{equation}
where $s_i$ with $i=x,y,z$ the Pauli matrices acting on the spin subspace, $\lambda_x$ and $\lambda_y$ the strength of a momentum-independent  spin-orbit coupling, and $\Upsilon$ a tensor that characterized an anisotropic Rashba-like term. The momentum independent SOC can be rewritten as a Kane-Mele Hamiltonian by performing a rotation by an angle $\varphi=\rm{arctan}(\lambda_y/\lambda_z)$ around the $x$-axis \cite{PRLGarcia2020}. We derived the model by imposing the 
P$2_1/$m symmetry of 1T'-WTe$_2$ by choosing a mirror axis along the $x$ direction $\mathcal{M}_x$, and two-fold rotation symmetry along the $z$-axis C$_2^z$. We consider also the possibility for a broken horizontal mirror plane symmetries which will lift also inversion symmetry. Such situation could arise due to the action of an electric field applied perpendicularly to the layer, the substrates, geometrical distortions (1T$_{\rm d}$ phase) or electrostatic gates, and is  characterized by the $\BB_z$ parameter. At the $Q$ points, the energy is given by $\varepsilon_{\alpha,\tau}^Q= \alpha \sqrt{\BB_x^2 + (\BB_z+\sigma \lambda)^2} $ where $\alpha=\pm 1$ for conduction and valence bands and $\sigma=\pm1$ for spins up and down and $\lambda = \lambda_x^2 + \lambda_y^2$. For small values of $\BB_z$, the spin-splitting is given by
\begin{equation}
 \Delta = 2 \lambda  \BB_z/\sqrt{\BB_x^2 + \lambda^2}.    \label{eq:spinsplit}
\end{equation}
Therefore, spin-splitting vanishes for $\BB_z=0$, in agreement with prior statements using symmetry arguments and microscopic calculations \cite{ShiLikun,PRLGarcia2020}. 
The spin texture is defined as the expectation value of the spin operator  $S_i({\bm{q}})\equiv \bra{\varepsilon(\bm{q})}s_{i}\ket{\varepsilon(\bm{q})}$ where $\ket{\varepsilon(\bm{q})}$ the eigenvectors of the system. At the $Q$ points, the spin texture is prescribed solely by the SOC parameters $\bm{S}(\bm{0})= (0, \sin(\varphi),\cos(\varphi))$. Therefore, measuring the spin-splitting provides a method to determine the strength of the SOC while the spin texture gives the canting angle. 

To determine the microscopic parameters, first-principles calculations were performed with the {\sc Siesta} \cite{Soler2002,Garcia2020} implementation of DFT, using GGA-PBE exchange-correlation functional, optimized norm-conserving pseudopotentials \cite{garcia2018_psml, Setten2018}, and a standard double-$\zeta$ polarized basis set \cite{Artacho1999}. 
SOC is included in the calculation using the fully-relativistic pseudopotential method proposed by Hemstreet {\it et al.} \cite{fr-pp-soc} and implemented in {\sc Siesta} by Cuadrado {\em et al.} \cite{Cuadrado2012,Garcia2020,CuadradoTBP}.
The electronic properties of monolayer WTe$_2$ converged for a 500 Ry real-space grid cutoff, a shifted 14$\times$8$\times$1 $k$-point mesh, and a Fermi-Dirac smearing of the electronic temperature of \emph{k}$_B$T = 8 meV. 
A vacuum thickness of more than 80 \AA~ was employed to avoid spurious interactions between the periodic replicae of the monolayer.
WTe$_2$ was fully relaxed in the 1T' phase using the conjugate gradient algorithm
until the forces on atoms were smaller than 0.01 eV/\AA ~ and the pressure on the cell less than 0.037 kBar. 
The optimized cell parameters (3.53{\AA} and 6.33{\AA}) are within 1\% of previous studies \cite{brown1966crystal}.

\begin{figure}
\includegraphics[width=0.48\textwidth]{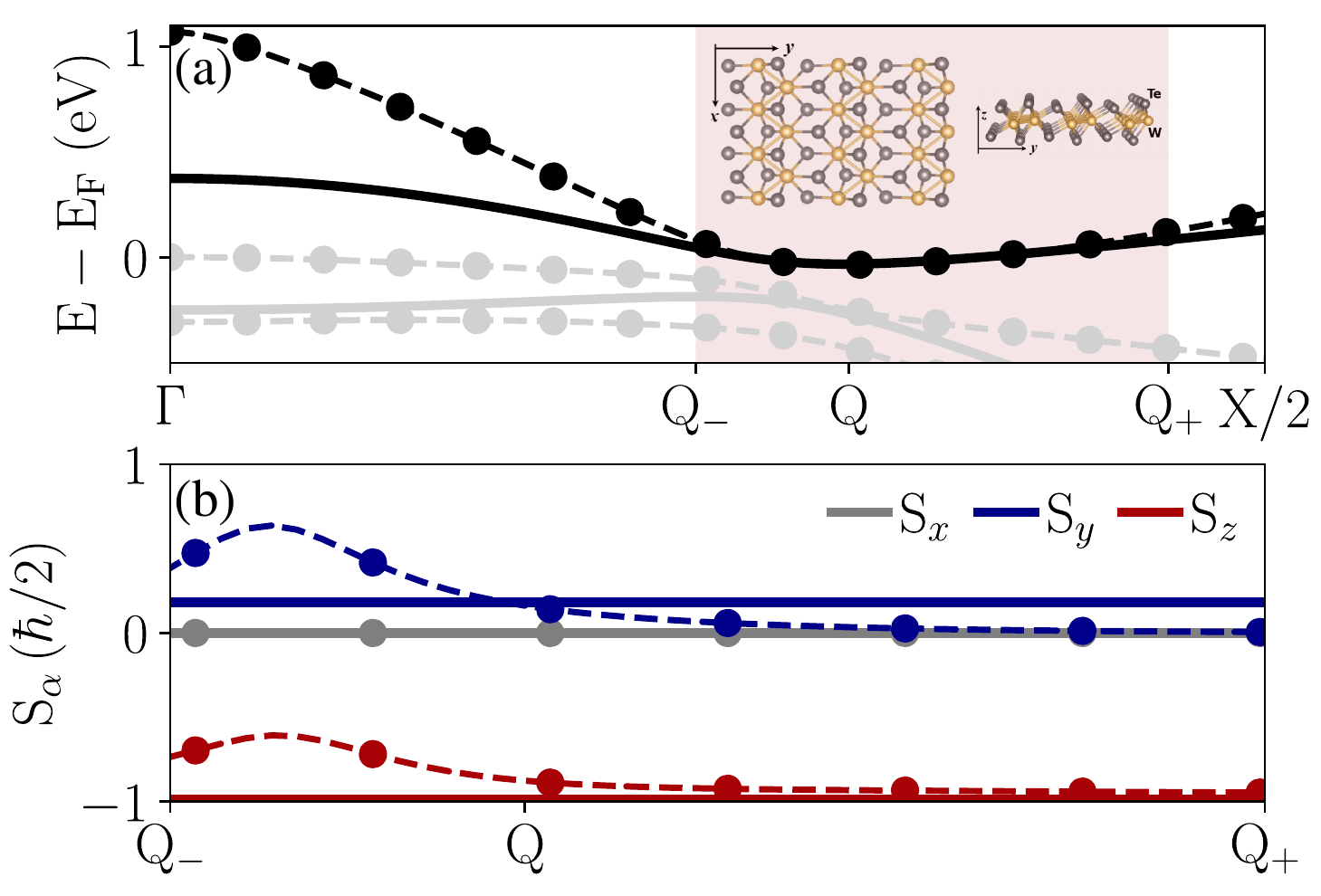}
\caption{(a) Band structure of 1T'-WTe$_2$ computed using DFT (dashed lines with symbols) and our model (solid lines) fitted considering the points in the shaded areas. The $Q$ points are defined by the minimum of the conduction band, highlighted in red. (b) Spin texture predicted by the DFT calculations and our fitted model. the structure of 1T'-WTe$_2$ is illustrated in inset (a). The definitions of $Q_\pm$ is provided in \cite{suppmat}
}
\label{fig_F1}
\end{figure}

Fig.\ \ref{fig_F1} shows the band-structure and spin-texture  computed using DFT (dashed lines with symbols) and our model (solid lines) for 1T'-WTe$_2$. Since our model was derived to describe the $Q$ points, we highlight in shaded red the range between two poinwts $Q_+$ and $Q_-$ that we use for the fits (technical details of the fits are given elsewhere \cite{suppmat}). As seen in  Fig.\ \ref{fig_F1}, the model reproduces well the main electronic and spin properties in the energy window of interest. From the band structure, we determined the Dirac velocity $v^{\rm F}_{x} = 0.7$ nm/fs, the anisotropy parameter $\zeta = 0.2$, the tilt velocity $v_{x}^{T}=0.4$ nm/fs, and the pseudo Zeeman field $\BB_{x}=0.7411$ eV. From the spin-texture, we identify the SOC strength $\lambda=0.188$ eV and a canting angle of $\varphi\approx 30^\circ$, in agreement with other theoretical and experimental calculations \cite{comment_angle,PRLGarcia2020,ZhaoCobden2020,ChengTan2020}. It is important to highlight that the DFT results predict a persistent spin texture, indicated by the constant plateaux spotted in the shaded region, also captured by our simplified model. 

\begin{figure}
\includegraphics[width=0.48\textwidth]{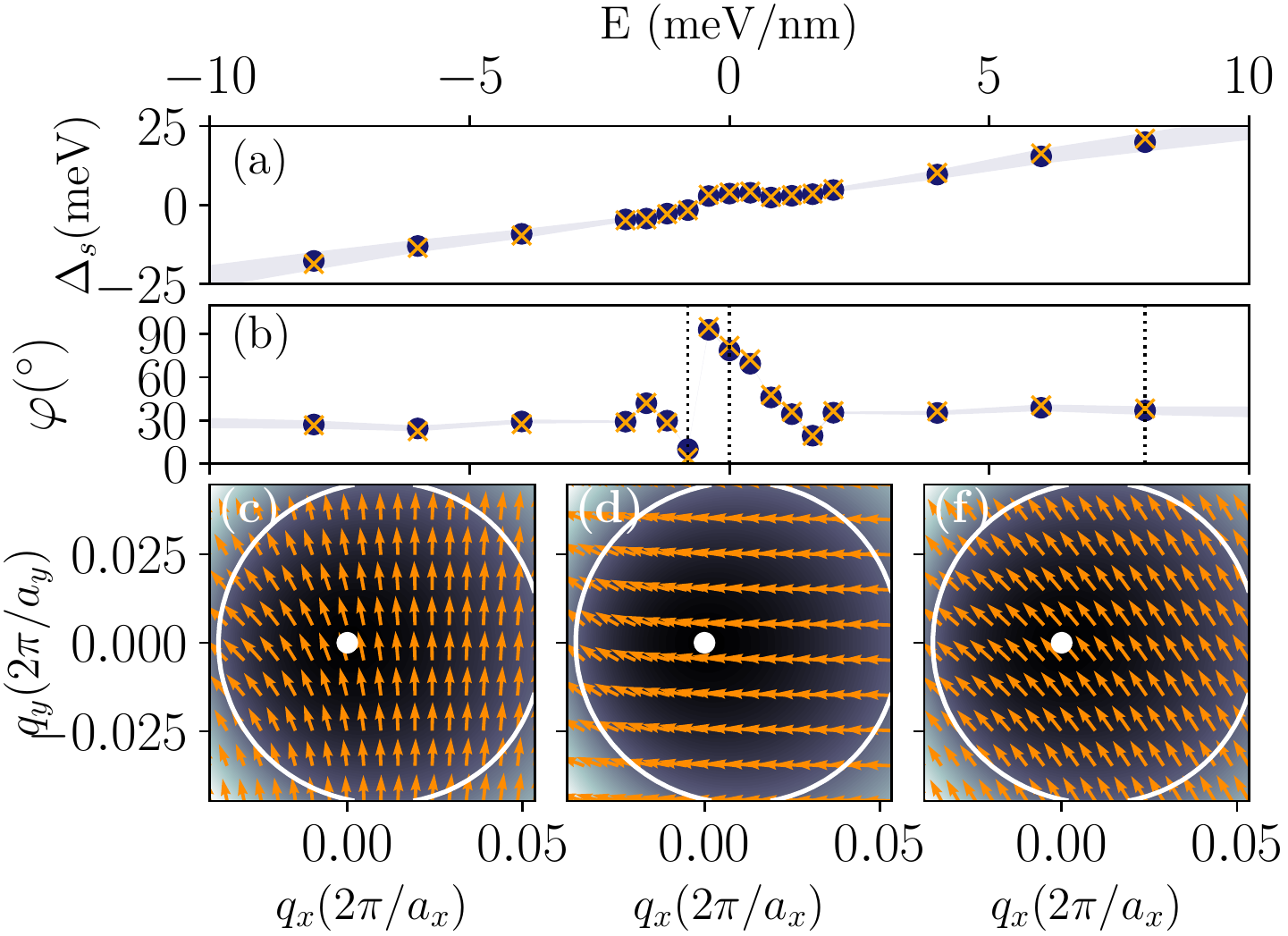}
\caption{(a) Comparison of the conduction band spin-splitting obtained {\it ab-initio} calculations (blue dots) and the prediction of our fitted model (orange crosses). (b) The canting angle of the spin texture computed via DFT compared to the canting angle computed from the spin-orbit coupling parameters of our model (orange crosses). To determine the canting angle and spin-splitting from DFT we average over a region of the Brillouin zone around the $Q$ points. The shaded region in (a) and (b) represents the standard deviation of the  splitting and canting angle around the Q points, and is a measurement of the fluctuations away from $Q$. (c-e) Examples of the regions we used to compute the spin-texture for different electric felds $E=$-0.8, 0.0 and 8 mV/nm. }
\label{fig_F2}\end{figure}

{\it Electric field effects.} To unveil the effect of external electric fields on the spin properties, we use DFT to compute the band structure and spin textures of 1T'-WTe$_2$ (relaxed at zero field) in presence of different electric fields ranging from -10 mV/nm to 10 mV/nm and fit the results to our model. As expected, all parameters remain insensitive to the field except $\mathcal{B}_z$ that breaks the inversion symmetry and the SOC parameters that define the canting angle $\varphi$. Moreover, the canting angle saturates at 30 degrees for electric field above 1 mV/nm, a value that matches previous experimental works. Fig.\ref{fig_F2}a shows the difference between the bands with opposite spins (orange crosses), as well as the parameter ${\rm \mathcal{B}}_z$ obtained from a fit to the DFT at each electric field (blue dots). The shaded region indicates the standard deviation of the splitting and canting angle around the $Q$ points, as a measure of their variations. We found a band splitting proportional to the applied electric field, while the spin splitting $\Delta$ depends linearly with the electric field. This splitting is due to the Stark effect and is consistent with previous previous researches\cite{Maximenko2022,Shi2019prb}. More importantly, our fitted $\BB_z$ evolves linearly with the electric field, and the ratio $\Delta/\BB_z \approx 0.49$ matches perfectly with Eq. \ref{eq:spinsplit} when using the fitted values for $\lambda$ and $\BB_x$.

Fig. \ref{fig_F2}b shows how the canting angle ($\varphi$) (related to the PST) evolves with the external electric field. It is clear that $\varphi$ is varying substantially  at low electric field, whereas it saturates to a critical angle $\varphi_c=30^o$ for high electric fields. These observations indicate two important things, (i) the PST is robust against electric field and its direction is electrically tunable, and (ii) our model captures this phenomenon through the renormalization of the SOC parameters. This is one of the central results of our article. The shaded regions displays the standard deviation, indicating a small discrepancy from the PST landscape. This is clearer from Fig. \ref{fig_F2}(c-e) where variations of the spin texture above the Fermi contour (white line) are seen to increase for higher electric fields. Such variations are attributable to the anisotropic Rashba field. 

{\it Electrically controlled chiral spin currents.} In WTe$_2$ the persistent spin texture defines the spin polarization direction in the QSH insulating regime \cite{PRLGarcia2020,ZhaoCobden2020,ChengTan2020}. To demonstrate that the electrically tunable PST enables full control of the spin polarization of spin currents (or canting angle), we determine the spin Hall conductivities using the Kubo-Bastin formula  \cite{Bastin1971, cresti2016rnc}:
\begin{align}\label{eq_Kubo}
\sigma_{ij}^{\alpha}=-2\hbar \Omega  \int_{-\infty}^{E_{\rm F}} dE\, \text{Im}\left( \text{Tr}\left[  \delta(E-{\cal H})J_{s,i}^{\alpha}\frac{dG^+}{dE}J_{j} \right]\right),
\end{align}
where $\Omega$ the area of the sample,  $J_{s,i}^\alpha \equiv \{J_{i},s_\alpha\}/2\,$ is the $i$-th component of the spin current density operator, with $\alpha=x,y,z$ denoting the spin polarization direction and $J_{j} \equiv (ie/\Omega\hbar) [ {\cal H} , R_{j}]$  the $j$-th component of the current density operator, with $e$ the electron charge and $R_{j}$ the position operator \cite{Fan2019}. The spectral operators $\delta(E-{\cal H})$ and $G^+\equiv 1/(E-{\cal H}+i0^+)$ are the Dirac delta and the retarded Green's function, respectively. We numerically computed the Kubo-Bastin formula using the kernel polynomial method \cite{Garcia2015, cresti2016rnc, Garcia2018, Fan2019} with 2000 Chebyshev expansion moments (equivalent to a 5\,meV of broadening), on a system containing 4 millions orbitals.

\begin{figure}
\includegraphics[width=0.48\textwidth]{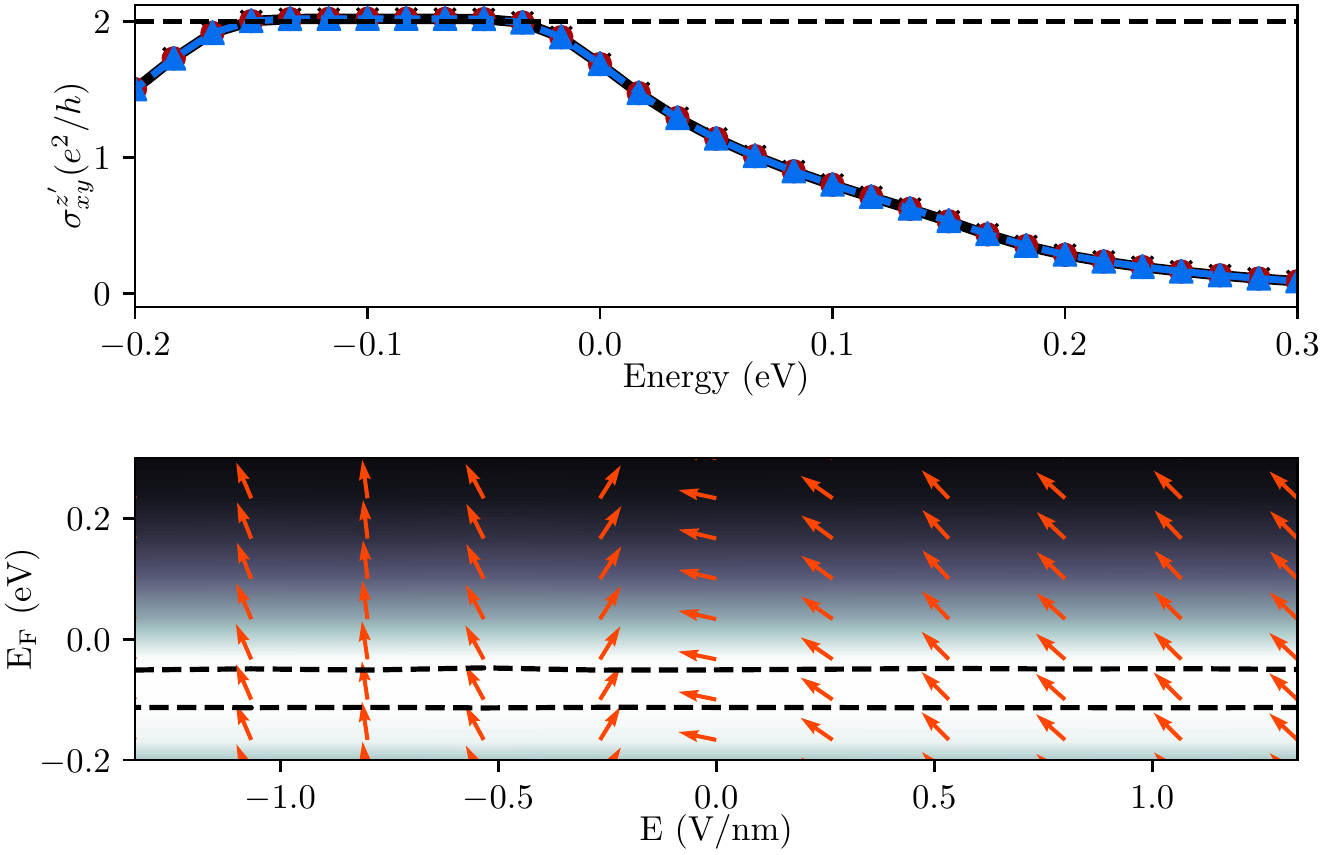}
\caption{(a) Spin Hall conductivity oriented along the $z'$ direction for $E=-0.8$ mV/nm (red), 0.0 mV/nm (black), and 8 mV/nm. (blue). The curves fell on top of each other demonstrating that these fields do not alter appreciable the topological properties of the system. (b) Spin polarization of the spin currents as a function of the electric field and Fermi level. The arrows indicate the direction of the spin polarization in the $yz$ plane. The dashed lines indicates the energy range of the conductivity plateau. 
}
\label{fig_F3}
\end{figure}

The results are presented in Fig. \ref{fig_F3}a, where the spin Hall conductivity is plotted along the $z'$ direction (prescribed by different canting angles depending on the electric field strength) for three different electric fields  $E=$-0.8, 0.0  and 8 mV/nm. Within the energy gap, the topological plateau remains pinned to its quantized value at $2e^2/h$ (three curves are almost superimposed), demonstrating that in this range of fields the small variations due to the anisotropic Rashba term do not jeopardize spin conservation and  topological protection. Fig. \ref{fig_F3}b shows the spin polarization of the spin current computed via the Kubo formula. The orange arrows indicates the direction of the spin on the $yz$ plane while the background color indicate its total magnitude. We observe a modulation of the spin direction with electric field that follows the spin texture behavior shown in Fig.\ref{fig_F2}b. Such trend holds for a large range of Fermi energies, enabling the control the spin polarization of either topological or bulk currents. For electric fields $E=-0.8$ mV/nm, the spin polarization lies along the $y$ direction, while at zero field it points toward the $z$-axis thus evidencing that a $90^\circ$ rotation for small and experimentally accessible fields is within the reach. 

The electrical modulation of the SOC parameters can be understood from the microscopic point of view by the formation of an electric dipole due to the distorted structure and presence of the electric field \cite{ShiLikun}. The resilience of the persistent spin texture derives from the robustness of the $Q$ points, that regardless to the strength of the electric field remains aligned with the $x$-axis, hence avoiding a detrimental effect of $q$-dependent spin-orbit terms. 

{\it Conclusions.} We have theoretically demonstrated that external electric fields below 10 mV/nm can tailor the spin polarization of spin Hall currents of 1T'-WTe$_2$ from in-plane to out-of-plane configuration. This behavior is a consequence controlled rotation of the persistent spin texture driven by the strength of the applied electric field, which ultimately dictates the direction of the spin polarization of the canted QSH effect. Importantly, such electric field control of the spin polarization of edge states does not compromise their topological robustness. These results promote 1T'-WTe$_2$ as an alternative source of tunable spin currents that could lead to all-electrical spintronic devices and open a door to explore other mechanism to tune the persistent spin texture such as strain or pressure. 
\begin{acknowledgments}
Supported by the EU H2020 Programme under Grants No. 881603 (Graphene Flagship), No 824140 (TOCHA, H2020-FETPROACT-01-2018) and No. 824143 (MaX Materials Design at the Exascale CoE) and by Spanish MICIU, AEI and EU FEDER (Grants No. PGC2018-096955-B-C43 and PCI2018-093120).  Z.Z. acknowledges support by the Ram\'on y Cajal program RYC-2016-19344 (MINECO/AEI/FSE, UE), the Netherlands sector plan program 2019-2023, Spanish MINECO (FIS2015-64886-C5-3-P). R.C. acknowledges the funding from the EU H2020 Programme under  the Marie Sklodoswka--Curie Grant No. 665919.
JY, PK, MGM and ZZ  acknowledge the computer resources at MareNostrum and the technical support provided by the Barcelona Supercomputing Center through Red Espa\~nola de Supercomputaci\'on (Grants No. RES-FI-2020-1-0018, RES-FI-2020-1-0014 and RES-FI-2020-2-0039).
ICN2 is funded by the Generalitat de Catalunya (CERCA Programme and Grants 2017SGR1506 and 2017SGR692), and is supported by the Severo Ochoa Centers of Excellence Program from Spanish MINECO under Grant No. SEV-2017-0706.
\end{acknowledgments}

\bibliography{bibwteQSHE}

\end{document}